\documentclass[a4paper,12pt]{article}
\usepackage{graphicx,rotating,hyperref,slashed,amsmath,xcolor,amssymb,amsfonts,colortbl,cite,subcaption,titlesec,empheq}
\usepackage[totalheight = 23cm, totalwidth = 17cm]{geometry}
\usepackage{physics}

\usepackage{listings,framed,xcolor,amsmath}
\colorlet{shadecolor}{gray!20}

\titleformat{\paragraph}
{\normalfont\normalsize\bfseries}{\theparagraph}{0.4em}{}
\titlespacing*{\paragraph}
{0pt}{1.25ex plus 0.2ex minus .2ex}{0.5ex plus .2ex}

\hypersetup{colorlinks, bookmarksopen, bookmarksnumbered,
citecolor=blue, linkcolor=blue, pdfstartview=FitH, urlcolor=blue}

\begin{document}

\title{\bf Global conformal symmetry in scalar-tensor theories}

\author{Eugeny Babichev$^{a}$, Christos Charmousis$^a$, Mokhtar Hassaine$^{b}$, Nicolas Lecoeur$^a$\\
\\[0.5cm]
\small{\em $^a$Universit\'e Paris-Saclay, CNRS/IN2P3, IJCLab, 91405 Orsay, France}\\
\small{\em $^b$Instituto de Matem\'atica, Universidad de Talca, Casilla 747, Talca, Chile}\\
}

\date{}   
{\let\newpage\relax\maketitle}

\begin{abstract}
We study a subclass of Horndeski gravity which has both global conformal and shift symmetries.
Global symmetries are characterised by the presence of a conserved current which has been shown to be of particular importance for the integrability features of the theory at hand yielding numerous compact object solutions. We find the general conserved current associated to global conformal symmetry of Horndeski theories. We discuss some of its properties, how it can be conveniently broken by physically relevant terms and, show how it is related to that of shift symmetry when shift symmetry is present. Given our results, we consider a particular theory and demonstrate how the presence of symmetries provides integrability for the given black hole solution. We then find the charged extension of the solution thanks to the conformal invariance of the Maxwell action.
\end{abstract}

\vspace{0.8cm}
{\em 
This article is part of a special issue of the International Journal of Modern Physics A, \\
which is devoted to the memory of V.A.Rubakov.}

\newpage
\section{Introduction}

Symmetries of a physical system are defined as the transformations that do not change the mathematical structure of the system. From Noether's theorem, linking symmetries to conservation laws, to the elegant mathematical framework of group theory, symmetries offer powerful tools for simplifying complex systems and predicting physical behavior. A striking example in this regard concerns Maxwell's equations, which were originally discovered  experimentally but could also have been derived from gauge symmetry. Indeed, one could even commence by defining the transformations and then finding the mathematical structure compatible with the transformations at hand. For example, the mathematical structure of relativity can be defined as the study of all structures compatible with the Lorentz symmetry. The determination of the symmetries of a system can also be a powerful instrument in order to generate nontrivial solutions from trivial ones or may play an important role for the integrability of the equations of motion. For all the above reasons amongst others, the identification of symmetries underlying a physical system  is not an academic question it is rather a fundamental one.

One of such fundamental symmetries in physics is conformal symmetry which plays an important  role in various theoretical setups as well as experimental observations. Indeed, conformal symmetry has attracted attention due to its rich mathematical structure and wide-ranging applications in condensed matter physics, high-energy physics, string theory, with a special mention on the AdS/CFT correspondence and holography. As mentioned before, symmetry can be helpful to obtain physically relevant solutions, and there is a simple example where conformal symmetry has played a crucial role in constructing the first counterexample to the no-hair conjecture. This concerns the case of a conformally coupled scalar field coupled to Einstein gravity in four dimensions~\cite{BBM, Be}. In this case,  the traceless energy-momentum tensor yields a geometric restriction (vanishing Ricci scalar, $R=0$), and this condition is of great utility for integrating the equations of motion, and also for establishing a uniqueness theorem~\cite{XZ}. In the presence of a cosmological constant and of the  self-interacting potential that does not break the conformal invariance of the scalar field action, the constraint $R=0$ is modified to $R=\mbox{cst}$, and analytic black hole solutions with (A)dS asymptotics can also be found in this case~\cite{Martinez:2002ru, Martinez:2005di}. One can be less restrictive and demand that conformal invariance be effective only for the scalar field equation and not necessarily for the scalar field action. It is from this point of view that Fernandes has shown recently that the standard conformal action of a scalar field in four dimensions can be modified by incorporating a scalar Gauss-Bonnet sector, and the resulting geometric constraint is now given by a constant linear combination of the Ricci scalar with the Gauss-Bonnet~\cite{Fernandes:2021dsb}. Note that the most general scalar field action in four dimensions yielding a second-order and conformally invariant equation has been identified in Ref.~\cite{Ayon-Beato:2023bzp}. We also have an interesting example where, starting from a quite general class of the Horndeski theory in $d=4$, and requiring the integrability of the equations, the resulting action was a mix of Lagrangians that had conformal invariance in four and five dimensions~\cite{Babichev:2023dhs}. In other words although the symmetry may not be present in a given dimension the actions at hand still maintain some of their integrability properties which makes them interesting beyond their field of direct application.

In the present work, we will dwell on this direction and consider scalar tensor theories admitting global symmetries. A widely encountered case in the literature is that of shift symmetry for the scalar field. This symmetry for the action has been proven to be incredibly useful for finding new solutions. The simple reason is that the scalar field equation can, for this subclass of symmetric theories, be written as a divergence of the Noether current related to the symmetry, and thus it can be directly integrated to give the first integral in special cases, such as, in particular, for spherically symmetric static cases.  This has given numerous black hole metrics starting from stealth static solutions~\cite{Babichev:2013cya} as well as stationary metrics~\cite{Charmousis:2019vnf,Anson:2020trg} to more recently black holes with primary hair~\cite{Bakopoulos:2023fmv} (for a very recent review of exact solutions see~\cite{Babichev:2023psy,Lecoeur:2024kwe}). Once black hole solutions are known one can also find physically relevant neutron star solutions~\cite{Charmousis:2021npl} and construct exotic wormhole geometries in closed form~\cite{Bakopoulos:2021liw}. It is then natural to seek theories employing a different symmetry that provides a corresponding conservation of a Noether current.  The symmetry we will consider is -- global conformal symmetry{\footnote{The combination of local conformal symmetry and global shift symmetry has been studied, but for bi-scalar theories, giving black holes with primary hair in certain cases~\cite{Charmousis:2014zaa}}} -- and we will show how to use these two symmetries in order to find spherically symmetric black hole solutions. 

In the next section we will discuss the relevant theories, sub-classes of Horndeski, with global conformal symmetry and then find the conserved current associated to the symmetry at hand. We will then discuss a particular example and how the symmetry is manifest in its integrability adding a Maxwell field (which is conformally invariant) for the latter part. We will also discuss the possible extensions and limitations of our findings.

\section{Scalar-tensor theories with global conformal symmetry}
The presence of a global conformal symmetry in the theory implies the existence of a Noether current which is accompanied by an equation for its conservation. In this section we introduce the sub-class of Horndeski theories possessing global conformal symmetry and find the Noether current corresponding to the symmetry at hand. We then also consider the presence of a term breaking the conformal symmetry in order to attend to more physically relevant theories and their solutions.

\subsection{Action of Horndeski theory with global conformal symmetry}
In four dimensions, the most general scalar-tensor theory with second order field equations is given by the Horndeski action~\cite{Horndeski:1974wa,Deffayet:2011gz,Kobayashi:2011nu},
\begin{equation}
\begin{split}
\label{action_H}
S\left[g_{\mu \nu}, \phi\right]=\int \mathrm{d}^4 x & \sqrt{-g}\left\{G_2-G_3 \square \phi+G_4 R+G_{4 X}\left[(\square \phi)^2-\phi_{\mu \nu} \phi^{\mu \nu}\right]\right. \\
& \left.+G_5 G^{\mu \nu} \phi_{\mu \nu}-\frac{G_{5 X}}{6}\left[(\square \phi)^3-3 \square \phi \phi_{\mu \nu} \phi^{\mu \nu}+2 \phi_{\mu \nu} \phi^{\nu \rho} \phi_\rho^\mu\right] \right\},
\end{split}
\end{equation}
where $G_i\equiv G_i(X,\phi)$ are arbitrary functions of the scalar field $\phi$ and the kinetic term $X=-\frac12(\nabla_\mu\phi\nabla^\mu\phi)$.

Within the Horndeski action there is a class of theories that admits global conformal invariance:
\begin{equation}
\label{scaling}
g_{\mu \nu} \rightarrow \omega^2 g_{\mu \nu}, \quad \phi \rightarrow \frac{\phi}{\omega},
\end{equation}
where $\omega$ is a constant scaling parameter. 
Indeed, the most general subclass of Horndeski theory possessing global conformal invariance with a nontrivial scalar and metric transformation  (\ref{scaling}) was given in Ref.~\cite{Padilla:2013jza}. In this class, the functions $G_i(X,\phi)$ have a specific dependence on $\phi$ and $X$:
\begin{equation}
\label{ai}
G_2 = \phi^4 a_2(Y),\quad G_3= \phi a_3(Y),\quad G_4 = \phi^2 a_4(Y) ,\quad G_5 = \frac{a_4(Y)}{\phi},
\end{equation}
where $a_i$ are arbitrary functions of $Y$ and $Y$ is defined as a combination of $X$ and $\phi$: $Y=X/\phi^4$. For further reference, we explicitly write the Horndeski action with global conformal symmetry:
\begin{equation}
\label{action_C}
    \begin{split}
S_C\left[g_{\mu \nu}, \phi\right]= & \int \mathrm{d}^4 x \sqrt{-g}\left\{\phi^4 a_2(Y)-\phi a_3(Y) \square \phi+\phi^2 a_4(Y) R
\right. \\
&  +\frac{a_{4 Y}} {\phi^2}\left[(\square \phi)^2-\phi_{\mu \nu} \phi^{\mu \nu}\right]
+\frac{a_5(Y)}{\phi} G^{\mu \nu} \phi_{\mu \nu}\\
&\left. -\frac{a_{5 Y}}{6 \phi^5}\left[(\square \phi)^3-3 \square \phi \phi_{\mu \nu} \phi^{\mu \nu}+2 \phi_{\mu \nu} \phi^{\nu \rho} \phi_\rho^\mu\right]\right\},
\end{split}
\end{equation}
where the subscript '$C$' refers to the global conformal symmetry of the action.
One can check that the action (\ref{action_C}) is invariant under transformation~(\ref{scaling}).

Our focus in this paper will be the study of (\ref{action_C}) however it is interesting to note that there are two more cases of global (conformal) symmetry that we can add to the generic one above. The first is still a global conformal transformation with a trivial transformation for the scalar field, namely
\begin{equation}
\label{scaling0}
g_{\mu \nu} \rightarrow \omega^2 g_{\mu \nu}, \quad \phi \rightarrow \phi,
\end{equation}
where the invariant action, in this case, is given by
\begin{eqnarray}
\label{action_H00}
S_{C0}=\int \mathrm{d}^4 x & \sqrt{-g}\left(a_2(\phi) X^2-a_3(\phi) X \square \phi+a_5(\phi)  G^{\mu \nu} \phi_{\mu}\phi_{\nu}+a_6(\phi)\, {\cal G}\right).
\end{eqnarray}
where ${\cal G}$ stands for the Gauss-Bonnet density. Note that the last term in~(\ref{action_H00}) is a part of the Horndeski action~(\ref{action_H}), see Ref.~\cite{Kobayashi:2011nu}.
It is interesting to note that for the particular values, $a_2=2\alpha, a_3=-4\alpha, a_5=4\alpha$ and $a_6=-\alpha \phi$, one recovers the action of Fernandes with linear Gauss-Bonnet coupling whose equation of motion has conformal symmetry while the action itself does not~\cite{Fernandes:2021dsb}.

We can also consider a trivial transformation of the metric instead, namely
\begin{equation}
\label{scaling00}
g_{\mu \nu} \rightarrow g_{\mu \nu}, \quad \phi \rightarrow \frac{\phi}{\omega}.
\end{equation}
Introducing $Z = X/\phi^2$, one can write the following invariant isometric action in this case
\begin{equation}
    \begin{split}
S_C\left[g_{\mu \nu}, \phi\right]= & \int \mathrm{d}^4 x \sqrt{-g}\left\{ a_2(Z)-\frac{1}\phi a_3(Z) \square \phi+ a_4(Z) R
\right. \\
&  +\frac{a_{4 Z}} {\phi^2}\left[(\square \phi)^2-\phi_{\mu \nu} \phi^{\mu \nu}\right]
+\frac{a_5(Z)}{\phi} G^{\mu \nu} \phi_{\mu \nu}\\
&\left. -\frac{a_{5 Z}}{6 \phi^3}\left[(\square \phi)^3-3 \square \phi \phi_{\mu \nu} \phi^{\mu \nu}+2 \phi_{\mu \nu} \phi^{\nu \rho} \phi_\rho^\mu\right]\right\}.
\end{split}
\end{equation}

\subsection{Noether current}
Given the action~(\ref{action_C}) with the symmetry~(\ref{scaling}), we need to figure out the Noether current corresponding to this symmetry. 
To do this we write the action in a general form,
\begin{equation}
\label{action_L}
S_C=\int \mathrm{d}^4 x\, L(g, \partial g, \partial \partial g, \phi, \partial \phi, \partial \partial \phi),
\end{equation}
where partial derivatives mean symbolically the partial derivative w.r.t. coordinates, e.g. $\partial\partial g$ imply $\partial_\mu\partial_\nu g_{\alpha\beta}$ etc.
The Euler-Lagrange equations read then,
\begin{equation}
\label{EOMs_L}
\begin{aligned}
0 &=\frac{\delta S_C}{\delta g^{\mu\nu}} =\frac{\partial L}{\partial g^{\mu \nu}}-\partial_\rho \frac{\partial L}{\partial\left(\partial_\rho g^{\mu \nu}\right)}+\partial_\rho \partial_\sigma \frac{\partial L}{\partial\left(\partial_\rho \partial_\sigma g^{\mu \nu}\right)}, \\
0 &=\frac{\delta S_C}{\delta \phi} =\frac{\partial L}{\partial \phi}-\partial_\rho \frac{\partial L}{\partial\left(\partial_\rho \phi\right)}+\partial_\rho \partial_\sigma \frac{\partial L}{\partial\left(\partial_\rho \partial_\sigma \phi\right)}.
\end{aligned}
\end{equation}
Let us collectively denote the metric and the scalar field as $\varphi^a$ for simplicity. By assumption, the action~(\ref{action_L}) is invariant under the transformation (in our case the global conformal transformation) $\varphi^a \rightarrow \varphi^a+\Delta \varphi^a$. This implies that $\delta L$ under this transformation is vanishing or equal to a total divergence $\delta L = \partial_\rho j^\rho$ (in the case where $j^\rho\not=0$, one refers to a quasi invariance, rather than to invariance). Using equations of motion~(\ref{EOMs_L}), one can write the change of $L$ as follows:
\begin{equation}
\delta L =\partial_\rho\left(\frac{\partial L}{\partial\left(\partial_\rho \varphi^a\right)} \Delta \varphi^a\right)+\partial_\rho\left(\frac{\partial L}{\partial\left(\partial_\rho \partial_\sigma \varphi^a\right)} \partial_\sigma \Delta \varphi^a-\Delta \varphi^a \partial_\sigma \frac{\partial L}{\partial\left(\partial_\rho \partial_\sigma \varphi^a\right)}\right),
\end{equation}
which implies that the conserved current $\mathfrak{J}_{\text {C}}^\rho$, such that $\partial_\rho \mathfrak{J}_{\text {C}}^\rho=0$, is given by,
\begin{equation}
\mathfrak{J}_{C}^\rho=\left(\frac{\partial L}{\partial\left(\partial_\rho \varphi^a\right)}-\partial_\sigma \frac{\partial L}{\partial\left(\partial_\rho \partial_\sigma \varphi^a\right)}\right) \Delta \varphi^a+\frac{\partial L}{\partial\left(\partial_\rho \partial_\sigma \varphi^a\right)} \partial_\sigma \Delta \varphi^a-j^\rho.\label{eq:def_J_frak}
\end{equation}
To be perfectly clear, the summation convention is used not only for the spacetime indices $\rho,\sigma$, etc., but also for the field indices $a$. For global conformal symmetry~(\ref{scaling}) it is convenient to present its transformation in the infinitesimal form $\omega=1-\epsilon$, so that $\Delta g^{\mu \nu}= 2 \epsilon g^{\mu \nu}$ and $\Delta \phi=\epsilon \phi$ for the field transformations.   In our case, the Lagrangian density $L$ itself, not just the action, is invariant under such a transformation, i. e. $j^\rho =0$.
Finally, a direct computation gives the following form\footnote{This comes from the fact that $L$ is a scalar density of weight $1$ and therefore $\mathfrak{J}_C^\rho$ is a vector density of weight $1$.} of~$\mathfrak{J}_{C}^\rho$:
\begin{equation}
\mathfrak{J}_{C}^\rho= \epsilon \sqrt{-g} \mathcal{J}_{C}^\rho,\label{eq:jfrakjcal}
\end{equation}
where $\mathcal{J}_{C}^\rho$ is a covariantly conserved Lorentz vector,
\begin{equation}
\label{Jconcerv}
\nabla_\rho \mathcal{J}_{C}^\rho=0 .
\end{equation}
The Noether current $\mathcal{J}_{C}^\rho$ can be presented as a sum of four terms (which corresponds to four free $a_i$ functions), 
\begin{equation}
\mathcal{J}_{C}^\rho=\sum_{i=2}^5 \mathcal{J}_{C,i}^\rho,
\end{equation}
where the $ \mathcal{J}_{C,i }^\rho$ are given in Appendix~\ref{app_A}.

Let us also make the assumption, that apart from global conformal invariance, the action is also shift symmetric, i. e.  the action is invariant under the change $\phi \to \phi +\mbox{cst}$. The scope of this is to have two conserved currents at the same time, which may imply (or not) an extra first integral of the equations of motion. We will see later, however, that this subclass of theories has its limitations. 

In case of presence of both global conformal and shift symmetries, the conformal current $ \mathcal{J}_{C,i}^\rho$ has a non-trivial relation to the Noether current of shift symmetry $J^\rho$
\begin{equation}
\label{JJ}
     \mathcal{J}_{C}^\rho = \phi J^\rho + B^\rho,
\end{equation}
where $J^\rho$ reads, 
\begin{equation}
\label{Jshift}
 J^\rho = \frac{1}{\sqrt{-g}}\frac{\delta S}{\delta \partial_\rho\phi},   
\end{equation}
and note that $B^\rho$ does not depend on $\phi$ explicitly.

The above relation~(\ref{JJ}) can be extended to the case of a theory possessing global conformal symmetry, but no shift symmetry. In this case one can still define the quantity $J^\rho$, although it is not a conserved quantity\footnote{Note that in the absence of shift symmetry there is an ambiguity in defining $J^\rho$, i.e. for equivalent actions that differ by a total derivative, $J^\rho$ given by~(\ref{Jshift}) do not coincide  in general. Working  with a fixed form of the action removes this uncertainty.}. The explicit expressions for $J^\rho$ and $B^\rho$ are given in Appendix~\ref{app_A}.

\subsection{Breaking symmetry by Einstein-Hilbert term}

As we discussed in the previous section, the conservation of the conformal current $\mathcal{J}_{\text {C}}^\rho$ is given by Eq.~(\ref{Jconcerv}). This equation, however, can be also seen as a consequence of the equations of motion. 
Indeed, denoting again the fields collectively as $\varphi^a$ for brevity, the transformation of $L$ under $\varphi^a\to\varphi^a+\Delta\varphi^a$ reads
\begin{equation}
    \delta L = \frac{\partial L}{\partial\varphi^a}\Delta \varphi^a+\frac{\partial L}{\partial\left(\partial_\rho\varphi^a\right)}\partial_\rho\Delta\varphi^a+\frac{\partial L}{\partial\left(\partial_\rho\partial_\sigma\varphi^a\right)}\partial_\rho\partial_\sigma\Delta\varphi^a.
\end{equation}
Combining now the Euler-Lagrange equations~(\ref{EOMs_L}), and the definition~(\ref{eq:def_J_frak}) of $\mathfrak{J}^\rho_C$ (where we recall that $j^\rho=0$), one obtains,
\begin{equation}
    \partial_\rho\mathfrak{J}^\rho_C=-\frac{\delta S_C}{\delta \varphi^a}\Delta\varphi^a+\delta L.
\end{equation}
Using the fact that $\delta L=0$ under the infinitesimal global conformal transformation $\Delta g^{\mu\nu}=2\epsilon g_{\mu\nu}$ and $\Delta\phi=\epsilon\phi$, together with the relation~(\ref{eq:jfrakjcal}) between $\mathfrak{J}^\rho_C$ and $\mathcal{J}^\rho_C$, one finally gets
\begin{equation}
    \nabla_\rho\mathcal{J}_C^\rho = -g^{\mu\nu}\mathcal{E}_{\mu\nu}^C-\phi\mathcal{E}^C_\phi,\label{JCEOMs}
\end{equation}
where for short, we have introduced the following notations,
\begin{equation}
    \mathcal{E}^C_{\mu\nu}\equiv \frac{2}{\sqrt{-g}}\frac{\delta S_C}{\delta g^{\mu\nu}},\quad \mathcal{E}^C_\phi\equiv \frac{1}{\sqrt{-g}}\frac{\delta S_C}{\delta \phi}.
\end{equation}
In a word, the conservation on shell of the global conformal Noether current $\mathcal{J}^\rho_C$ can be immediately understood from Eq.~(\ref{JCEOMs}).
\\

Let us now consider adding a symmetry breaking term to a theory with global conformal symmetry. This may be important, for example, if we want to have an Einstein-Hilbert term in the action, so that we can restore General relativity in some limit. Clearly, the action with a minimally coupled Einstein-Hilbert term is not invariant under global conformal transformations. This in turn implies that the current $\mathcal{J}_C^\mu$ is no longer conserved. Let us see how Eq.~(\ref{Jconcerv}) is modified when we add a general symmetry breaking term in the action, i.e. we consider the action $S = S_C+S_b$, where $S_C$ is an invariant action under~(\ref{scaling}) and $S_b$ breaks the global conformal invariance. 
One gets that, on shell,
\begin{equation}
\label{JCEOMs-}
    \nabla_\mu \mathcal{J}_C^\mu = g^{\mu\nu}\mathcal{E}^b_{\mu\nu}+ \phi\mathcal{E}_\phi^b,
\end{equation}
where we used the equations of motion $\mathcal{E}^C_{\mu\nu}+\mathcal{E}^b_{\mu\nu}=0$, $\mathcal{E}^C_{\mu\nu}+\mathcal{E}^b_{\mu\nu}=0$ together with~Eq.~(\ref{JCEOMs}).

Take for instance $S_b$ to be the pure Einstein-Hilbert action, with Lagrangian $L_b=\sqrt{-g} \zeta R$, where the constant $\zeta$ is normally identified with Planck mass squared.  In this case, $\mathcal{E}_{\mu\nu}^b=2\zeta G_{\mu\nu}$ while $\mathcal{E}_\phi^b=0$, and hence~(\ref{JCEOMs-}) becomes simply,
\begin{equation}
\label{JC_R}
    \nabla_\mu \mathcal{J}_{\text {C}}^\mu =-2\zeta R.
\end{equation}
The conformal current is thus not conserved, and is rather sourced by the symmetry breaking term, the Ricci scalar.

\section{Asymptotically flat black hole solutions and its extension}

We are now ready to apply our findings for particular examples of black hole solutions in theories with (partially broken) global conformal symmetry. 
We will focus our attention on the manifestation of the symmetry within the solutions, and discuss possible implications of the symmetry for finding new solutions. Our first example is a class of asymptotically flat black hole solutions in Horndeski theory, presented in Ref.~\cite{Babichev:2017guv}. In this reference, it has been noticed that a class of Horndeski theories giving rise to exact black hole solutions has (partial) global conformal invariance, but no consequence of the presence of the symmetry and the properties of the solution were discussed. 

Here we take another look at this black hole solutions from the perspective of the discussion above, i.e. the presence the conformal current and its (partial) conservation.  The lesson drawn from this example will allow us to further generalise the solution of Ref.~\cite{Babichev:2017guv} to include the Maxwell gauge field which is  (locally) conformally invariant.

\subsection{Asymptotically flat solutions in Horndeski theory from the perspective of symmetries of the theory}
\label{sec:BCL}
A class of asymptotically flat black hole solutions presented in Ref.~\cite{Babichev:2017guv} was found in a subclass of Horndeski theory,
\begin{equation}
\label{BCL action}
\begin{split}
S_1  = & \int \mathrm{d}^4 x \sqrt{-g}\left\{\left[\zeta+\beta \sqrt{(\partial \phi)^2 / 2}\right] R-\frac{\eta}{2}(\partial \phi)^2\right.\\
&\left. -\frac{\beta}{\sqrt{2(\partial \phi)^2}}\left[(\square \phi)^2-\left(\nabla_\mu \nabla_\nu \phi\right)^2\right]\right\} .
\end{split}
\end{equation}
where $\zeta=M^2_\text{Pl}/(16\pi)$ and $\eta$ and $\beta$ are dimensionless constants (note that the scalar field has dimension of mass). In terms of the general action~(\ref{action_H}), the theory~(\ref{BCL action}) is given as 
$G_2 = \eta X, \; G_4 = \zeta +\beta \sqrt{-X},\;$ and  $G_3=G_5=0$. The theory is shift-symmetric, i.e. the action is invariant under the change $\phi\to \phi+\mbox{cst}$. Note also that for $\zeta=0$, the theory also enjoys global conformal invariance~(\ref{scaling}). The pure Einstein-Hilbert term $\zeta R$ in the action breaks this symmetry. 
It is, however, necessary to include the Einstein-Hilbert term, since without it the theory with $G_4 = \beta \sqrt{-X}$ does not propagate the spin-2 graviton degrees of freedom, which renders the theory physically uninteresting (see for example~\cite{Kobayashi:2011nu}). Moreover, the solution of~\cite{Babichev:2017guv} does not have a well-defined limit as $\zeta\to 0$. 

Let us analyse possible solutions of the theory having in mind symmetries of the action~(\ref{BCL action}). The exact shift symmetry implies conservation of the shift current,
\begin{equation}
\label{J_conserv}
    \nabla_\mu J^\mu =0.
\end{equation}
The partially broken global conformal symmetry, on the other hand, yields a non-conservation of the conformal current $\mathcal{J}_{\text {C}}^\mu$, as given in~(\ref{JC_R}).
Assuming a static spherically symmetric ansatz,
\begin{equation}
\begin{aligned}
\mathrm{d} s^2 & =-h(r) \mathrm{d} t^2+\frac{\mathrm{d} r^2}{f(r)}+r^2\left(\mathrm{d} \theta^2+\sin ^2 \theta \mathrm{d} \varphi^2\right), \\
\phi & =\phi(r),
\end{aligned}
\end{equation}
one obtains the following general solution for $\phi'$ by integrating~(\ref{J_conserv}):
\begin{equation}
\label{solphi}
    \phi' = \pm\frac{\beta}{\eta r^2}\sqrt{\frac{2}{f}} - \frac{C}{\eta r^2}\frac{1}{\sqrt{fh}},
\end{equation}
where $C$ is a constant of integration, standing for scalar charge (associated with shift symmetry). 

In~\cite{Babichev:2017guv}, it has been assumed that the scalar charge is zero, based on the assumption that the norm of the shift-symmetry current $J^2=(J^\mu J_\mu)$ should be finite everywhere, including the horizon of a black hole. It has been noted in~\cite{Babichev:2016rlq} that, however, the norm $J^2$ diverges at the horizon of a black hole in a theory with linear scalar-Gauss-Bonnet coupling $\phi \mathcal{G}$ for a solution that seems physically reasonable. Therefore one may think of relaxing the condition of finiteness of the norm $J^2$ and to consider arbitrary $C$ in (\ref{solphi}). Nevertheless, one arrives to the same conclusion, $C=0$, asking a more physically motivated requirement of finiteness of $X$ (see a discussion on this subject in~\cite{Babichev:2024txe}). Hence, the condition $C=0$ is in fact equivalent to $J^\mu=0$.
 
Having established that $J^\mu = 0$, we now move on to analyse the conformal current equation~(\ref{JC_R}). Using~(\ref{JJ_relation})-(\ref{Bi}) it is easy to see that for the theory~(\ref{BCL action}) we have
\begin{equation}
    \label{JJ_BCL}
    \mathcal{J}_C^\rho = \phi\, J^\rho.
\end{equation}
Substituting the above relation into (\ref{JC_R}) and taking into account that $J^\mu=0$ we immediately get 
\begin{equation}
\label{R0}
    R=0.
\end{equation}
In Ref.~\cite{Babichev:2017guv} this condition of vanishing Ricci scalar was identified by combining equations of motion, without referring to the (partial) global conformal symmetry. The use of the global conformal symmetry by the application of the global current conservation~(\ref{JC_R}) leads immediately to the conclusion that the spacetime has vanishing Ricci scalar.

After finding~(\ref{solphi}) with $C=0$ and (\ref{R0}), it is then a matter of simple calculations to use other equations of motion to find the metric,
\begin{equation}
\label{solfh}
    f(r)=h(r)=1-\frac{\mu}{r}-\frac{\beta^2}{2 \zeta \eta r^2},
\end{equation}
where $\mu$ is another integration constant. Note that the presence of the term proportional to $1/r^2$ is a consequence of global conformal invariance (see below with the introduction of the conformally invariant Maxwell term (\ref{solfhe}). Substituting the above expression in~(\ref{solphi}), one can explicitly calculate $\phi=\phi(r)$. The corresponding expression can be found in Ref.~\cite{Babichev:2017guv}.

Before closing this section, it is instructive to consider briefly one more possibility which is allowed for our action due to global shift symmetry of the scalar; namely a linear time dependence for the scalar field, $\phi=q t +\psi(r)$, see~\cite{Babichev:2010kj,Babichev:2013cya,Babichev:2006vx}. Then one can show that the scalar charge $C$ is always forced to be zero from the presence of the $(tr)$ equation of motion~\cite{Babichev:2015rva}. In a certain sense the integration constant $q$ takes over the presence of the scalar charge $C$. Therefore this imposes that $J^r=0$. However note that $J^t\neq 0$ and this gives for our conformal current,
\begin{equation}
    \mathcal{J}_C^t = q t \, J^t
\end{equation}
which in turn (due to the explicit presence of $t$) requires that the Ricci scalar is no longer zero, $R\neq 0$ according to (\ref{JC_R}). The corresponding solution was found in~\cite{Babichev:2017guv} to be non asymptotically flat. We see here that time dependence in $\phi$ sources in turn the Einstein Hilbert term present in the action precisely due to the presence of the additional conformal symmetry. We therefore set $q=0$ for the forthcoming section.  

\subsection{Extension to include the Maxwell field}
Now, having in mind the example of black hole solutions we considered above, where the role of (partial) conformal symmetry was identified,  one may think of possible extensions of this solution.  Indeed, since our focus in the paper is the conformal symmetry, it is natural to include the Maxwell  electromagnetic field $A_\mu$ in the discussion. The action for standard electromagnetic field  has local conformal invariance, hence also global invariance. To be precise, we consider a scalar-tensor-vector theory, with the (partial) global conformal invariance whose action on the dynamical fields reads
\begin{equation}
\label{scaling2}
g_{\mu \nu} \rightarrow \omega^2 g_{\mu \nu}, \quad \phi \rightarrow \frac{\phi}{\omega}, \quad A_\mu \to A_\mu.
\end{equation}
Clearly, under the transformation~(\ref{scaling2}) the Maxwell action,
\begin{equation}
    \label{EM}
    S_{EM} = \int \mathrm{d}^4 x \sqrt{-g}\left(-\frac14 F_{\mu\nu}F^{\mu\nu}\right),
\end{equation}
where $F_{\mu\nu} = \partial_{\mu}A_\nu -\partial_{\nu}A_\mu$, is invariant. We may guess therefore that the combination of the two actions,
\begin{equation}
\label{action_e}
    S_{Ext} = S_1 + S_{EM},
\end{equation}
also has a solution, similar to the one found in Ref.~\cite{Babichev:2017guv}.

Following the logic of the previous discussion in Sec.~\ref{sec:BCL}, we find that the theory~(\ref{action_e}) has shift symmetry $\phi\to\phi +\mbox{cst}$, partial global conformal symmetry and, in addition, the gauge symmetry of the vector field. It is not difficult to see that the Maxwell term does not contribute to both currents -- associated to shift and conformal symmetry, -- therefore those currents for the theory~(\ref{action_e}) are the same as for the original theory~(\ref{BCL action}). 

This implies that the expression for the scalar field in terms of metric~(\ref{solphi}) remains the same, with $C=0$ for the reasons explained in~\ref{sec:BCL}. Similarly, the conclusion that the Ricci scalar is zero~(\ref{R0}) is not altered by the presence of the the Maxwell term. On the other hand, the variation of the action~(\ref{action_e}) with respect to $A_\mu$ gives Maxwell's equations, 
\begin{equation}
    \label{Maxwell}
    \nabla_{\mu}F^{\mu\nu}=0.
\end{equation}
As we assumed spherical symmetry and staticity, one can look to a purely electric ansatz $A_{\mu}= \left\{ A_0(r),0,0,0\right\}$. It is easy to  integrate~(\ref{Maxwell}) as 
\begin{equation}
    \label{solA}
    A_0'(r) = -\sqrt{\frac{h}{f}}\frac{Q}{r^2},
\end{equation}
where $Q$ is a constant of integration that has a meaning of an electric charge of a black hole. 

For the moment, we have obtained both $\phi$ and $A_\mu$ in terms of the metric function, and in addition we have a constraint $R=0$.  Taking into account the form of the metric in the case considered above, Eq.~(\ref{solfh}), and the Reissner–Nordstr\"om solution in general relativity, one can guess the solution for $f$ and $h$ in the theory~(\ref{action_e})
\begin{equation}
    \label{solfhe}
    f(r)=h(r)=1-\frac{\mu}{r}+ \frac{Q^2}{4\zeta r^2}-\frac{\beta^2}{2 \zeta \eta r^2}.
\end{equation}
In the above expression we combined linearly the 'usual' term from the backreaction of the Maxwell field and the gravity modification term, since they both come with the same power of $r$. 
Direct substitution of the Eq.~(\ref{solphi}) with $C=0$, Eqs.~(\ref{solA}) and~(\ref{solfhe}) in the metric equations verifies that this is indeed a black hole solution. 

To sum up, an electrically charged black hole solution to the theory with the action~(\ref{action_e}) is given by the metric~(\ref{solfhe}) and the scalar and gauge fields,
\begin{equation}
    \label{sole}
      \phi' = \pm\frac{\beta}{\eta r^2}\sqrt{\frac{2}{f}},\qquad A_\mu dx^{\mu}= \frac{Q}r dt,
\end{equation}
where $f$ is given in~(\ref{solfhe}). The dyonic extension of this purely electric solution is straightforward, and given by
\begin{eqnarray}
&&f(r)=h(r)=1-\frac{\mu}{r}+ \frac{Q^2+Q_m^2}{4\zeta r^2}-\frac{\beta^2}{2 \zeta \eta r^2}, \qquad  \phi' = \pm\frac{\beta}{\eta r^2}\sqrt{\frac{2}{f}},\nonumber\\
&&A_\mu dx^{\mu}= \frac{Q}r dt+Q_m\cos\theta d\varphi,
\end{eqnarray}
where $Q_m$ stands for  the magnetic charge. 

\section{Discussion}

In this paper we have studied the consequences of global symmetries to scalar tensor theories and how they affect the integrability of the equations of motion. 
We examined global conformal symmetry and we have given the relevant conserved current and the equations of motion. The conformal current has a neat 
relation to the shift symmetric current, when the latter is also present.

It is well known that shift symmetric Horndeski theories admit a plethora of black hole solutions (see for example the recent reviews~\cite{Babichev:2023psy,Lecoeur:2024kwe}) as well as solutions with primary scalar hair~\cite{Bakopoulos:2023fmv,Baake:2023zsq,Bakopoulos:2023sdm}). 
Here we considered a theory with partially broken global conformal symmetry as well as shift symmetry and studied in detail what the symmetries imply on the solution itself. We found the electrically charged (and dyonic) extension of the black hole solution presented in Ref.~\cite{Babichev:2017guv}. 
We also noted how the combined presence of both symmetries can drastically restrict possible solutions in the presence of a linear time dependent scalar.
Thus the  presence of combined symmetries can be both beneficial or disadvantageous for finding new solutions, depending on a specific set up.

For future work, it would be interesting to study the extension of our example construction to higher dimensions. 
In this case,  one could consider breaking  global conformal invariance via higher order Lovelock terms in a similar fashion as was considered for local conformal invariance recently~\cite{Fernandes:2021dsb}. 
Another interesting possibility would be to extend our study to four dimensional beyond Horndeski theories with exact (unbroken) global conformal and shift invariance, since the requirement of both exact symmetries in Horndeski action leads to non-dynamical spin-2 perturbations.

But without doubt the most interesting question would be to study theories with (broken or unbroken) global conformal invariance which are not shift symmetric and thus have an explicit $\phi$ dependence. The additional difficulty in this case resides with the fact that the current depends not only on the (first or second) derivatives of the scalar but also on the scalar itself. This makes the first step towards solving the equations of motion technically harder. For example, in the case of shift symmetry, one obtains an algebraic condition on the first derivative of the scalar with respect to the metric (see for example~\cite{Babichev:2013cya}). The presence of symmetry and the conserved current may nevertheless be important in making progress towards obtaining black hole vacua in such non shift symmetric theories. 

\section*{Acknowledgments}
We thank Adolfo Cisterna for interesting discussions. E.B. and C.C. acknowledge the support of ANR grant StronG (ANR-22-CE31-0015-01). The work of M.H. is partially supported by FONDECYT grant
1210889. 

\appendix
\section{Noether current for Horndeski theories with global conformal symmetry}
\label{app_A}
Here we give expressions for each contribution of $G_i$ to the Noether current of for Horndeski theories with global conformal symmetry:
\begin{equation}
\label{JCi}
\begin{aligned}
\mathcal{J}_{C,2}^\rho& =  - \phi G_{2 X} \partial^\rho \phi, \\
\mathcal{J}_{C,3}^\rho& =  G_3 \partial^\rho \phi+\phi\left[\left(G_{3 X} \square \phi+G_{3 \phi}\right) \partial^\rho \phi-G_{3 X} \partial_\alpha \phi \nabla^\alpha \nabla^\rho \phi\right] \\
\mathcal{J}_{C,4}^\rho& =  -6 G_{4 \phi} \partial^\rho \phi+\phi\left\{2 G_{4 X} G^{\rho \nu} \partial_\nu \phi+2 G_{4 X \phi}\left(\partial_\mu \phi \nabla^\mu \nabla^\rho \phi-\square \phi \partial^\rho \phi\right)\right. \\
& +G_{4 X X}\left(2 \square \phi \partial_\mu \phi \nabla^\mu \nabla^\rho \phi-2 \partial^\mu \phi \nabla_\mu \nabla_\nu \phi \nabla^\nu \nabla^\rho \phi\right. \\
& \left.\left.-\left[(\square \phi)^2-\left(\phi_{\mu \nu}\right)^2\right] \partial^\rho \phi\right)\right\} \\
\mathcal{J}_{C, 5}^\rho& =  G_{5 X}\left\{-4 X G^{\rho \mu} \phi_\mu+\square \phi \phi_\mu \phi^{\mu \rho}-\phi^\mu \phi_{\mu \nu} \phi^{\nu \rho}-\frac{1}{2}\left[(\square \phi)^2-\phi_{\mu \nu} \phi^{\mu \nu}\right] \phi^\rho\right\} \\
& +2 G_{5 \phi}\left(\square \phi \phi^\rho-\phi_\mu \phi^{\mu \rho}\right)+\phi\left\{-\mathcal{L}_{5 X} \phi^\rho-2 G_{5 \phi} G^{\rho \mu} \phi_\mu-G_{5 X}\left[G^{\rho \mu} X_\mu\right.\right. \\
& \left.+R^{\rho \mu} \square \phi \phi_\mu-R_{\mu \nu} \phi^\mu \phi^{\nu \rho}-R^{\alpha \rho \beta \mu} \phi_\mu \phi_{\alpha \beta}\right]+G_{5 \phi X}\left[\frac{\phi^\rho}{2}\left((\square \phi)^2-\phi_{\mu \nu} \phi^{\mu \nu}\right)\right. \\
& \left.+\square \phi X^\rho-X_\mu \phi^{\mu \rho}\right]+G_{5 X X}\left[\frac{X^\rho}{2}\left((\square \phi)^2-\phi_{\mu \nu} \phi^{\mu \nu}\right)\right. \\
& \left.\left.-X_\mu\left(\square \phi \phi^{\mu \rho}-\phi^{\alpha \mu} \phi_\alpha^\rho\right)\right]\right\}.
\end{aligned}
\end{equation}
where we used the notation $X_\mu = \partial_\mu X$. Note that here $G_i$ are not arbitrary, but are constrained to have the form~(\ref{ai}) in order for the action to be invariant under transformations~(\ref{scaling}).

The above expressions for $\mathcal{J}_{C,i}^\rho$ can be written in more compact form:
\begin{equation}
\label{JJ_relation}
    \mathcal{J}_{C, i}^\rho = \phi J^\rho_i +B^\rho_i,
\end{equation}
where 
\begin{equation}
\label{Ji}
\begin{aligned}
J_{2}^\rho& =  - G_{2 X} \partial^\rho\phi , \\
J_{3}^\rho& = \left(G_{3 X} \square \phi+G_{3 \phi}\right) \partial^\rho \phi-G_{3 X} \partial_\alpha \phi \nabla^\alpha \nabla^\rho \phi \\
J_{4}^\rho& = 2 G_{4 X} G^{\rho \nu} \partial_\nu \phi+2 G_{4 X \phi}\left(\partial_\mu \phi \nabla^\mu \nabla^\rho \phi-\square \phi \partial^\rho \phi\right)\\
& +G_{4 X X}\left(2 \square \phi \partial_\mu \phi \nabla^\mu \nabla^\rho \phi-2 \partial^\mu \phi \nabla_\mu \nabla_\nu \phi \nabla^\nu \nabla^\rho \phi\right. \\
& \left.-\left[(\square \phi)^2-\left(\phi_{\mu \nu}\right)^2\right] \partial^\rho \phi\right) \\
J_{5}^\rho& = 
 -\mathcal{L}_{5 X} \phi^\rho-2 G_{5 \phi} G^{\rho \mu} \phi_\mu-G_{5 X}\left[G^{\rho \mu} X_\mu\right. \\
& \left.+R^{\rho \mu} \square \phi \phi_\mu-R_{\mu \nu} \phi^\mu \phi^{\nu \rho}-R^{\alpha \rho \beta \mu} \phi_\mu \phi_{\alpha \beta}\right]+G_{5 \phi X}\left[\frac{\phi^\rho}{2}\left((\square \phi)^2-\phi_{\mu \nu} \phi^{\mu \nu}\right)\right. \\
& \left.+\square \phi X^\rho-X_\mu \phi^{\mu \rho}\right]+G_{5 X X}\left[\frac{X^\rho}{2}\left((\square \phi)^2-\phi_{\mu \nu} \phi^{\mu \nu}\right)\right. \\
&\left.-X_\mu\left(\square \phi \phi^{\mu \rho}-\phi^{\alpha \mu} \phi_\alpha^\rho\right)\right] ,
\end{aligned}
\end{equation}
and 
\begin{equation}
\label{Bi}
\begin{split}
B_2^\rho & =  0, \\
B_{3}^\rho & =  G_3 \partial^\rho \phi, \\
B_{4}^\rho & =  -6 G_{4 \phi} \partial^\rho \phi, \\
B_{5}^\rho& =  G_{5 X}\left\{-4 X G^{\rho \mu} \phi_\mu+\square \phi \phi_\mu \phi^{\mu \rho}-\phi^\mu \phi_{\mu \nu} \phi^{\nu \rho}-\frac{1}{2}\left[(\square \phi)^2-\phi_{\mu \nu} \phi^{\mu \nu}\right] \phi^\rho\right\} \\
& +2 G_{5 \phi}\left(\square \phi \phi^\rho-\phi_\mu \phi^{\mu \rho}\right).
\end{split}
\end{equation}
The expressions~(\ref{Ji}), valid even for non-shift symmetric theories, coincide with those introduced in Ref.~\cite{Kobayashi:2011nu}. When shift symmetry is absent, $J^\rho$ is not a conserved current. If a theory enjoys shift symmetry, i.e. when all the functions $G_i$ do not depend explicitly on $\phi$, then $J^\rho=\frac{1}{\sqrt{-g}}\frac{\delta S}{\delta \partial_\rho\phi}$ is a conserved Noether current, $\nabla_\mu J^\mu = 0$. The latter conservation law  coincides with the equation of motion of the scalar field.

\bibliographystyle{utphys}
\bibliography{bibconfsym.bib}

\end{document}